\documentstyle[prd,aps,preprint,psfig]{revtex}
\newcommand{\del}{\partial}

\newcommand{\x}{{\mathbf{x}}}

\newcommand{\n}{{\mathbf{n}}}
\newcommand{\k}{{\mathbf{k}}}

\newcommand{\ti}{\textit}
\newcommand{\f}{\frac}

\newcommand{\bb}{\bibitem}
\newcommand{\BF}{\begin{figure}\begin{center}}
\newcommand{\EF}{\end{center}\end{figure}}
\newcommand{\BE}{\begin{equation}}
\newcommand{\EE}{\end{equation}}
\newcommand{\BEA}{\begin{eqnarray}}
\newcommand{\EEA}{\end{eqnarray}}
\begin{document}
%@twocolumnfalse\endcsname
%%
%%
\tighten
\draft

\title{COBE Constraints on a
Compact Toroidal Low-density Universe}		
\author{Kaiki Taro Inoue}
\address{Yukawa Institute for Theoretical Physics, Kyoto University,
Kyoto 606-8502, Japan}
\date{\today}

\maketitle
\begin{abstract}

In this paper, the cosmic microwave background (CMB) 
anisotropy in a multiply-connected compact flat 3-torus model 
with the cosmological constant is investigated.
Using the COBE-DMR 4-year data, a full Bayesian analysis 
revealed that the constraint on the topology of the flat 3-torus 
model with low-matter-density is less stringent.
As in compact hyperbolic models, 
the large-angle temperature fluctuations can be
produced as the gravitational potential decays at
the $\Lambda$-dominant epoch well after the last scattering.
The maximum allowed number $N$ of images of the cell (fundamental domain) 
within the observable region at present is approximately 49 for
$\Omega_m=0.1$ and $\Omega_\Lambda=0.9$ whereas $N\sim8$
for $\Omega_m=1.0$ and $\Omega_\Lambda=0$. 
\end{abstract}

%%%%%%%%%%%%%%%% INTRODUCTION %%%%%%%%%%%%%%%%%%%
\section{INTRODUCTION}

For a long time, cosmologists have assumed the simply connectivity
of the spatial hypersurface of the universe. If it is the case, 
the topology of closed 3-spaces is limited to that of a 3-sphere 
if Poincar\'{e}'s conjecture is correct.  
However, if we assume that the spatial
hypersurface is multiply connected, then the geometry of 
spatially finite models can be flat or hyperbolic as well.
The metric describes only the local geometry. 
We should be able to
observe the imprint of the ``finiteness'' of the spatial geometry if
it is multiply connected on scales of the order of the 
particle horizon or less, in other words, if we live in a 
``small universe''. 
\\
\indent
For flat multiply connected models without the cosmological constant,
various constraints using the COBE DMR data have been obtained
\cite{Sokolov,Starobinsky,Stevens,Oliveira1,Oliveira2,Levin1,Roukema}.
Assuming that the initial power spectrum is scale-invariant ($n\!=\!1$)
the suppression of the temperature fluctuations on scales beyond the 
typical size of the cell $L$\footnote{$L$ can be defined as
twice the diameter which is defined as the maximum of the minimum geodesic
distance between two points over the space.}
leads to a decrease in the large-angle power.  
For 3-torus models without the cosmological constant 
in which the cell (fundamental domain) is a cube, 
the constraint is $L>0.8 R_\ast$ where $R_\ast$ is the 
comoving radius of the last scattering surface(horizon radius)
\cite{Sokolov,Starobinsky,Stevens,Oliveira1}. 
It should be emphasised that 
the constraint itself \ti{does not imply that the ``small universe'' 
is ruled out} since the possible maximum expected number
of the copies of the cell within the observable region at present
is approximately 8. For models in which one side of the cell 
is longer than the others, the constraint for the smallest
topological identification scale(=the diameter of the smallest 
ball which can wrap around the space) 
can be less stringent\cite{Roukema}. 
\\
\indent
On the other hand, recent observations of distant
supernova Ia \cite{Perlmutter,Riess} imply
the existence of ``missing energy `` which possesses 
a negative pressure and equation-of-state ($w\equiv p/\rho$)
in the form of either a cosmological constant, vacuum energy density or
a slowly-varying spatially inhomogeneous component ``quintessence''.
Together with the first observation of the height and the position
of the first Doppler peak by {{\sc Boomerang}}\cite{Boomerang} and  
{{\sc MAXIMA}} \cite{MAXIMA} and the COBE-DMR data, the constraint for
the cosmological constant is 
$0.69<\Omega_\Lambda<0.82$ and 
for the matter(cold dark matter, baryon plus radiation) 
$0.28<\Omega_m<0.42$ assuming adiabatic initial perturbations\cite{Balbi}.
\\
\indent
For low-matter-density
models with flat geometry a bulk of large-angle cosmic microwave
background(CMB) fluctuations 
can be produced as the gravitational 
potential decays at the $\Lambda$-dominant epoch 
$1+z\sim (\Omega_\Lambda/\Omega_0)^{1/3}$.
Recent works \cite{Aurich,Inoue2,CS,AS} 
have shown that the angular power spectrum $C_{l}$ 
is completely consistent with
the COBE-DMR data for some compact hyperbolic models
which are incompatible with the previous analyses \cite{Bond1}.
Because the angular sizes of fluctuations produced at the late epoch 
are large compared to those on the last scattering 
for flat or hyperbolic geometry, we expect that
the constraints for compact flat models
with low-matter-density can be also significantly loosened.
\\
\indent 
In this paper, the CMB 
anisotropy in a multiply-connected compact flat 3-torus model 
with or without the cosmological constant is investigated.
In sec II we briefly describe the time evolution of the scalar 
perturbation in the locally flat Friedmann-Robertson-Walker 
models which will be used for computing the
CMB anisotropy. In sec III we compare the angular power spectrum
in a flat 3-torus model with the cosmological constant to the one 
in the ``standard'' 3-torus model with $\Omega_{tot}\!=\!1$.  
In sec IV a full Bayesian analysis using the COBE-DMR
data has been carried out for giving the constraint on the 
minimum size of the cell. 
\\
\indent
%%%%%%%%%%%%%%%% SECTION II %%%%%%%%%%%%%%%%%%%
\section{CMB Anisotropy}
\indent
In what follows we assume that the matter consists of two
components:a relativistic one and a non-relativistic one
($\Omega_m=\Omega_r+\Omega_n$). First of all we consider the
evolution of the background space. 
From the Friedmann equation, the
integral representation for the time evolution of the 
scale factor $a$ (normalised to 1 at present time) in terms of the 
conformal time $\eta$ takes the form
\BE
\eta(a)=H_{0}^{-1}\int_0^a 
\f{da}{\sqrt{\Omega_{n0}a+\Omega_{r0}+\Omega_\Lambda a^4}}
\label{eq:a}
\EE
where $H_0$, $\Omega_{r0}$ and $\Omega_{n0}$ are the 
Hubble parameter, the density parameter for relativistic and 
non-relativistic matter at present, respectively.
Note that (\ref{eq:a}) can be written in terms of the elliptic
integral of the first kind which is too complex to describe here. 
The radius of the last scattering surface $R_\ast$ in comoving coordinates
can be obtained by integrating $\eta$ from $1/(z_{LSS})\sim 1/1100$ to
$1$. One can easily see that the presence of 
$\Omega_\Lambda$ makes $R_\ast$ to a larger value. For a fixed
$H_{0}$ and the total density at present $\Omega_{tot}$, the presence
of a cosmological constant or vacuum energy implies a
decrease in the cosmological expansion rate in earlier
epoch which augments the age of the universe and the 
size of the causally connected region.  
As for the radiation density, 
the standard scenario of the thermal history gives the present density 
$\Omega_{r0}=\Omega(photon)+\Omega(neutrino)=
4.15\times10^{-5}h^{-2}$ assuming three generations for neutrinos 
\cite{Kolb-Turner}.
Applying the value to (\ref{eq:a}) one obtains 
$R_\ast=4.98$ for $(\Omega_{n0}, \Omega_\Lambda)=(0.1,0.9)$
while $R_\ast=1.99$ for $(\Omega_{n0}, \Omega_\Lambda)=(1.0, 0)$. 
\\
\indent
Next, we consider the time evolution of the scalar-type perturbation. 
Assuming that the
anisotropic pressure is negligible, in the Newtonian gauge, the Newtonian
curvature perturbation $\Phi$ in $k$ space satisfies 
\cite{Mukhanov},
\BE
\Phi''+3 {\cal{H}}(1+c_s^2)\Phi'+c_s^2k^2\Phi+
(2 {\cal{H}}'+(1+3 c_s^2){\cal{H}}^2)\Phi=-4 \pi G a^2 p \Gamma,
\label{eq:scalar}
\EE
\BEA
c_s^2&=&\f{1}{3} \f{\rho_r}{\rho_n+\f{3}{4}\rho_r},
\nonumber
\\
\Gamma&=&\f{\rho_n}{\rho_r+\f{3}{4}\rho_n}\Biggl(
\f{3}{4}\f{\delta \rho_r}{\rho_r}-\f{\delta \rho_n}{\rho_n}
\Biggr ),
\EEA
where ' denotes the conformal time derivative $d/d\eta$, 
${\cal{H}}\equiv a'/a$, G is the Newton's constant, p is the total 
pressure, $c_s$ is the sound speed of the
fluid and $\Gamma$ corresponds to the entropy perturbation defined as
$p \Gamma\equiv \delta p - \f{p'}{\rho '}\delta \rho$. Assuming  
a pure adiabatic perturbation ($\Gamma=0$), the amplitudes of $\Phi$
for the non-decaying mode are constant in the radiation dominant epoch
if $k \eta_{eq}<<1$\cite{Mukhanov}. 
Hence the non-decaying mode of the perturbation that enters the Hubble
radius after the radiation-matter equality time $\eta_{eq}$ can be 
obtained by setting the initial condition $\Phi'(0)=0$. 
In practice, we approximate the solution of (\ref{eq:a})
by interpolating polynomial functions which are used for numerically
solving the second order 
ordinary differential equation (\ref{eq:scalar}).
\\
\indent 
For superhorizon perturbations, the amplitudes of $\Phi$ are constant 
in the radiation dominant epoch ($\Phi/\Phi(0)=1$) and in
the matter dominant epoch ($\Phi/\Phi(0)=9/10$) 
However, in the $\Lambda$-dominant
epoch $\Phi$ gradually decays as $1/a$ that will play a critical
role in characterising the large-angle temperature fluctuations
for $\Lambda$-models. 
\\
\indent
Assuming a pure adiabatic initial condition and neglecting
the anisotropic pressure, the temperature 
fluctuation on large angular scales can be 
written in terms of the curvature perturbation
$\Phi$ as 
\BE
\f{\Delta T}{T}(\n)=-\f{1}{3} \Phi(\eta_\ast, (\eta_0-\eta_\ast)\n)
-2\int _{\eta_\ast}^{\eta_0} \f {\del \Phi(\eta, (\eta_0-\eta)\n)}
{\del \eta} d \eta,
\label{eq:SW}
\EE
where $\n$ denotes the unit vector which points towards the sky
and $\eta_\ast$ and $\eta_0$ correspond to the last scattering 
and the present conformal time, respectively\cite{Sachs,HSS}. 
The first term in the right-hand 
side in (\ref{eq:SW}) describes the ordinary Sachs-Wolfe (OSW)
effect while the second term
describes the integrated Sachs-Wolfe (ISW) effect. 
Note that the Sachs-Wolfe
formula (\ref{eq:SW}) is valid only for perturbations whose scale is 
sufficiently larger than the horizon radius at the last scattering
$k<1/\eta_\ast$. 
From (\ref{eq:a}), one obtains the conditions $k<20.5 H_0$
for $(\Omega_m, \Omega_\Lambda)=(1.0, 0)$ and $k<9.9 H_0$
for $(\Omega_m, \Omega_\Lambda)=(0.1,0.9)$. The effect of 
the acoustic oscillations becomes important for fluctuations 
on smaller scales. However, for large angular scales ($l<15$), 
contributions from acoustic oscillations are negligible. 
%%%%%%%%%%%%%%%% SECTION III %%%%%%%%%%%%%%%%%%%
\section{Power spectrum}
Let us now consider cosmological models whose spatial geometry is
represented by a flat 3-torus which is obtained by gluing 
the opposite faces of a cube by three translations. Then the wave numbers
of the square-integrable eigenmodes of the Laplacian are restricted to
the discrete values $k_i=2 \pi n_i/L$, ($i=1,2,3$) where $n_i$'s run
over all integers. The angular power spectrum is written as
\BEA
C_l&=&\sum_{\k\ne 0} \f{8 \pi^3 {\cal{P}}_\Phi(k)F_{kl}^2}{k^3 L^3},
\nonumber
\\
F_{kl}&=&\f{1}{3}
\Phi(\eta_\ast) j_l(k(\eta_o\!-\!\eta_\ast))
\!+2 \int_{\eta_\ast}^
{\eta_o}\!\!\!\!\!\!d \eta\, 
\f{d\Phi}{d \eta}j_l(k(\eta_o\!-\!\eta)),
\label{eq:psT3}
\EEA
where ${\cal{P}}_\Phi(k)$ is the initial power spectrum for $\Phi$
and $k\equiv\sqrt{k_1^2+k_2^2+k_3^2}$. From now on we assume
the scale-invariant 
Harrison-Zeldovich spectrum (${\cal{P}}_\Phi(k)
\!=\!const.$) as the initial power spectrum.  
\BF
\centerline{\psfig{figure=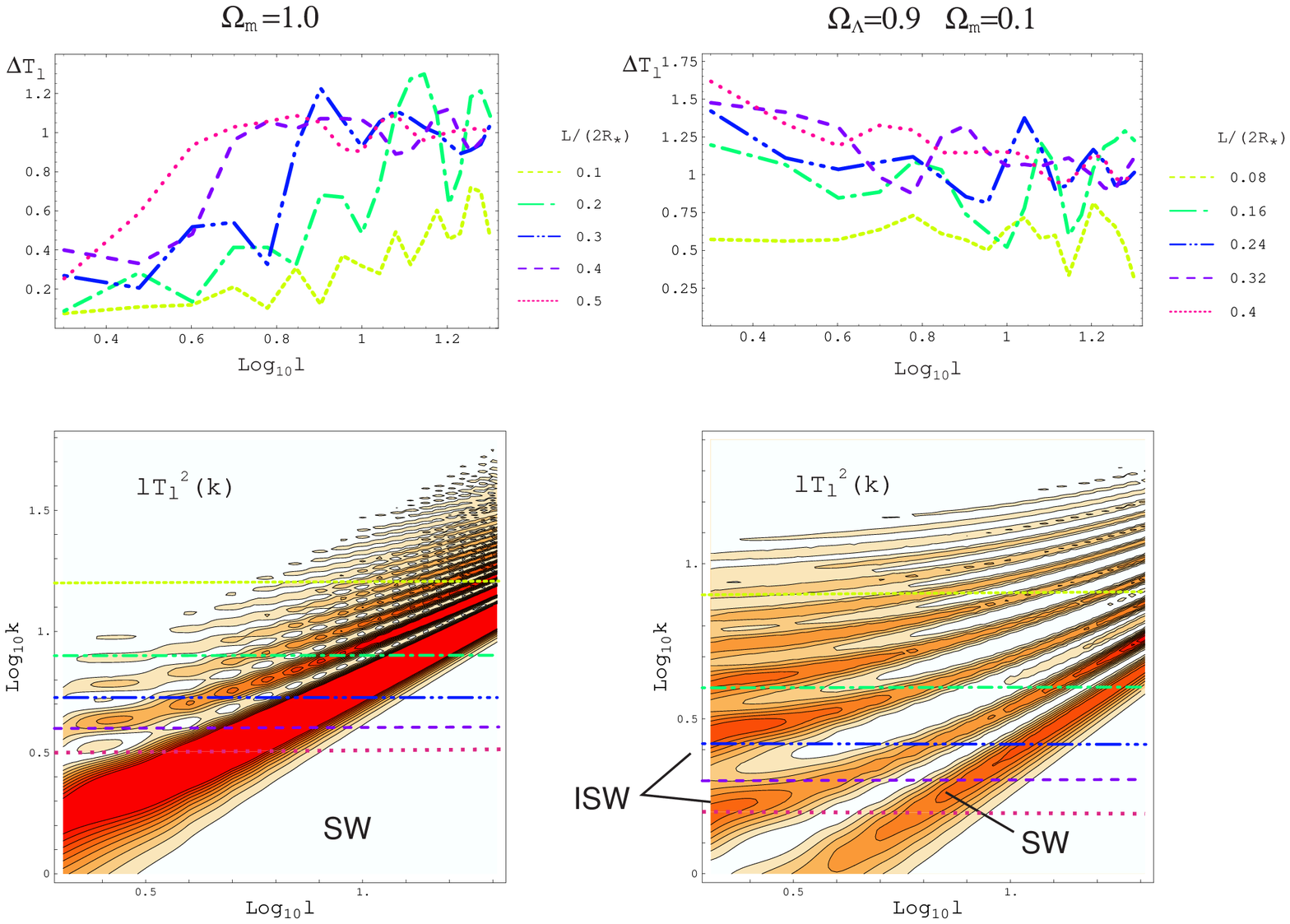,width=19cm}}
\caption{Suppression in large-angle power for 3-torus models with or
without the cosmological constant. 
A large-angle suppression in $\Delta T_l\equiv \sqrt{l(l+1)C_l/(2
\pi)}$ (all the plotted values are 
normalised by $\Delta T_{20}$ with infinite volume) 
occurs for the ``standard'' 3-torus model 
$(\Omega_m,\Omega_\Lambda)\!=\!(1.0,0)$
at $l\!<\!l_{cut}\!\sim\! 2\pi R_\ast/L-1$ 
while such a prominent suppression is
\ti{not} observed for the 3-torus model with 
($\Omega_m, \Omega_\Lambda$)=(0.1,0.9). 
The transfer function $T_l(k)$ describes how each $k$-mode
contributes to the power for a particular angular scale $l$. 
The corresponding first eigenvalues $k_1$ are plotted
as horizontal lines(lower figures).
The unit of $k$ is $H_0$. }
\label{fig:TL}
\EF
Now let us estimate the angular scale below which the power spectrum
is suppressed owing to the mode-cutoff $k_{cut}=2 \pi/L$. 
First of all we consider the transfer function $T_l(k)$
which is defined by
\BE
\f{2l+1}{4 \pi}C_l= \int T_l^2(k){\cal P}(k) \f{dk}{k},
\EE
that describes how the spatial information is included in the angular 
power. For a given $k$, the contribution to the angular power
on smallest angular scales comes from the OSW effect
where the transfer function can be written in terms of the 
spherical Bessel function as $T^{OSW}_l(k)=j_l(k\eta_0)$. 
Because $j_l(x)$'s have the first peaks at $x\sim1+l$, the 
angular cutoff $l_{cut}$ is determined by $l_{cut}=2 \pi \eta_0/L-1$.
For instance, $l_{cut}\sim 5$ for $L=\eta_0=2 H_0^{-1}$. 
Thus the large-angle suppression scale is determined by the
largest fluctuation scale at the last scattering. On smaller angular scales 
$l>l_{cut}$, the second peak in the power corresponds to the
fluctuation scale of the second eigenmode at the last scattering. 
This behavior is analogous to the acoustic oscillation where the
oscillation scale is determined by the sound horizon at the last scattering.
On angular scales larger than $l_{cut}$ 
one observes another oscillation feature
in the power for models with a smaller cell 
where $k_{cut} \eta_0$ is sufficiently
large which is apparently determined by the behavior of the 
first eigenmode. From the asymptotic form 
\BE
j_l(x)\sim \f{1}{x} \sin(x-l \pi/2),~~~x>>\f{l(l+1)}{2},
\EE
one notices that the angular scale of the oscillation 
for asymptotic values $k_{cut} \eta_0>>l(l+1)/2$ is $\Delta l=2$. 
For intermediate values of $k_{cut}\eta_0$, $\Delta l$ take much 
larger values (see figure \ref{fig:TL}).
\\
\indent
So far, we have studied the effect of the non-trivial topology 
on the OSW contribution only which is sufficient for constraining 
the topology of the ``standard'' 3-torus model with $\Omega_{tot}=1$.
However, for low-matter-density models, 
one cannot ignore the (late) ISW contribution which is
generated by the decay of the gravitational potential at the 
$\Lambda$-dominant epoch. The crucial point is that 
they are produced at \ti{later} time $1+z\sim
(\Omega_\Lambda/\Omega_0)^{1/3}$ well after the last scattering.
Although it is impossible to generate 
fluctuations beyond the size of the cell (in 3-dimensional sense),
that does not necessarily mean that any   
fluctuations on large angular scales (in 2-dimensional sense)
cannot be produced.
Suppose that a fluctuation is produced at a point nearer to us,
then the corresponding angular scale 
becomes large if the background geometry 
is flat or hyperbolic. 
Therefore, one can expect that the
suppression on large angular scales owing to the mode-cutoff
is not stringent for low-matter-density models. 
As shown in 
figure \ref{fig:TL}, the angular powers for a model with
$(\Omega_\Lambda, \Omega_m)\!=\!(0.9,0.1)$ are almost flat.
In contrast to the ``standard'' model, the transfer function
for low-matter-density models 
distributes in a broad range of $k$ 
that implies the additional late production of the fluctuations
which contribute to the angular power. 
Surprisingly, in the low-matter-density models with small volume,  
the slight excess power due to the ISW effect is cancelled out
by the moderate suppression owing to the mode-cutoff which leads to
a flat spectrum. However, as observed in the ``standard'' 3-torus 
model, the power spectra have prominent oscillating features. The 
oscillation scale for $l<l_{cut}$ is 
determined by the first eigenmode. The peaks in the angular power  
correspond to the first SW ridge and the first and the second 
and other ISW ridges. Is such an oscillating feature 
already ruled out by the current observation?
We will see the results of our Bayesian analyses for testing 
the goodness-of-fit to the COBE data in the next section.
%%%%%%%%%%%%%%%% SECTION IV %%%%%%%%%%%%%%%%%%%
\section{Bayesian analysis}
\indent
In general, the covariance in the temperature  
at pixel $i$ and pixel $j$ in the sky map can be written as
\BE
M_{ij}=<T_i T_j>=\sum_{l}<a_{lm}a_{l'm'}>W_l W_{l'}Y_{l m}(\hat{n_i})
Y_{l' m'}(\hat{n_j})+<N_i N_j>
\label{eq:M}
\EE
where  $a_{lm}$ is an expansion coefficient with respect to a
spherical harmonic $Y_{lm}$, $<>$ denotes an ensemble
average taken over all initial conditions, positions and
orientations of the observer, 
$T_i$ represents the temperature at pixel $i$, $W_l^2$  is the
experimental window function that includes the effect of 
beam-smoothing and finite pixel size,
$\hat{n_i}$ denotes the unit vector towards the center of pixel $i$
and $<N_i N_j>$ represents the noise covariance between 
pixel $i$ and pixel $j$. 
If the temperature fluctuations form a 
homogeneous and isotropic random Gaussian
field then the covariance matrix can be written 
in terms of the power spectrum $C_l$ as
\BE
M_{ij}=\f{1}{4 \pi}\sum_{l}(2l +1) W_l^2 C_l P_l
(\hat{n_i}\cdot \hat{n_j})+<N_i N_j>
\label{eq:MI}
\EE
where $P_l$ is the Legendre function. 
Then the probability distribution function of 
the pixel temperature $\vec{T}$ for the Gaussian field is 
\BE
f(\vec{T}|C_l)=\f{1}{(2 \pi)^{N/2} \det^{1/2}M(C_l) } 
\exp \Biggl (\f{1}{2}\vec{T}^T\cdot M^{-1}(C_l) \cdot \vec{T}\Biggr),
\EE
where $N$ is the number of pixels. Bayes's theorem states that
the probability distribution function of a set of parameters
$\vec{A}$ given the data $\vec{T}$ is
\BE
f(\vec{A}|\vec{T})\propto f(\vec{T}|\vec{A}) f(\vec{A}).
\EE
If we assume a uniform prior distribution, \ti{i.e.} 
taking $f(\vec{A})$ to be constant,
the probability distribution function of a power spectrum $C_l$
is then 
\BE
\Lambda(C_l|\vec{T})\propto\f{1}{\det^{1/2}M(C_l) } 
\exp \Biggl (\f{1}{2}\vec{T}^T\cdot M^{-1}(C_l) \cdot \vec{T}\Biggr). 
\label{eq:LL}
\EE
\\
\indent
Before applying the method to the flat 3-torus models 
one must be aware that a flat 3-torus we are considering is globally
anisotropic although it is globally homogeneous.
In contrast to the standard infinite models, the 
fluctuations form an anisotropic Gaussian field for a fixed orientation
if the initial fluctuation is Gaussian. 
In other words, for a given $l$, a set of 
$a_{lm}$'s ($-l \! \le \!m\!\le\! l$) are not $2l+1$ independent 
random numbers. In order to see this, we write a plane wave 
in terms of eigenmodes in the spherical coordinates,
\BE
e^{i \k\cdot\x}=\sum_{l m}  b_{k l m}~ j_l(k x) Y_{lm}(\n), ~~~
b_{k l m}=4 \pi (i)^l~ Y^\ast_{l m} (\hat \k),
\EE
where $\hat \k$ denotes the unit vector in the direction of $\k$.
Then we have
\BE
a_{lm}=\sum_\k \Phi(k)~~b_{k l m} F_{k l}
\EE
where $ F_{k l}$ is given by (\ref{eq:psT3}). Because
$a_{lm}$ is linear in $\Phi(k)$, it is Gaussian. However,
they are not independent since 
$b_{k l m}$'s are proportional to 
spherical harmonics. The statistical 
isotropy is recovered iff
the size $L$ of the cube becomes infinite where $\hat \k$ takes
the whole values for a given $k$. If one marginalises the
likelihood with respect to the $SO(3)$ transformation, or
equivalently, the orientation of the observer,
the distribution function of $a_{lm}$ becomes non-Gaussian
since $a_{lm}$ is written in terms of a sum of a product of 
a Gaussian variable times a variable that is determined by 
a spherical harmonic $Y_{lm}(\hat \k)$ where $\hat \k$ 
is a random variable which obeys a uniform distribution.
Thus in principle, in order to compare a whole set of 
fluctuation patterns over the isotropic ensemble one should
use the likelihood function which is different
from the Gaussian one. 
\\
\indent
Nevertheless, 
we first carry out a Bayesian analysis using the Gaussian likelihood
function (\ref{eq:LL}) which depends only on $C_l$'s (which is 
$SO(3)$ invariant) 
in order to estimate the effect of the mode-cutoff
imposed by the periodic boundary conditions. 
In the following analysis, we use the inverse-noise-variance-weighted
average map of the 53A,53B,90A and 90B  
COBE-DMR channels. To remove the emission from the galactic
plane, we use the extended
galactic cut (in galactic coordinates) \cite{Banday}.
After the galactic cut, best-fit monopole and dipole are removed  
using the least-square method.
To achieve efficient analysis in computation,  we  
compress the data at ``resolution 6'' $(2.6^o)^2$  
pixels  into one at ``resolution 5'' $(5.2^o)^2$ 
pixels for which there are 1536 pixels in the
celestial sphere and 924 pixels surviving the extended galactic cut.
The window function is given by $W_l=G_l F_l$ where $F_l$ are the Legendre
coefficients for the DMR beam pattern \cite{Lineweaver} and $G_l$ 
are the Legendre
coefficients for a circular top-hat function with area equal to the
pixel area which account for the pixel smoothing effect 
(which is necessary for ``resolution 5'' pixels since the 
COBE-DMR beam FWHM is comparable to the pixel size) \cite{Hinshaw}.
To account for the fact that we do not have useful information
about monopole and dipole anisotropy, the likelihood must be
integrated over $C_0$ and $C_1$ in principle. However, in practice  
we set $C_0=C_1=100 mK^2$ which renders the likelihood insensitive to
monopole and dipole moments of several $mK$.  We also assume
that the noise in the pixels is uncorrelated from pixel to pixel 
which is found to be a good approximation\cite{Tegmark-Bunn}.
\\
\indent
From figure \ref{fig:LLres5F} one can see that the constraint
on the size of the cell is less stringent for the low-matter-density
model as expected from the shape of the power spectrum. 
The effect of the suppression of the large-angle power is
not prominent unless the cell size is sufficiently smaller
than the observable region ($L<0.8H_o^{-1}=0.08\times 2R_\ast$ 
for $\Omega_m=0.1$). However, the likelihood function varies rapidly 
as the size increases since the power is jagged in $l$.
Unfortunately, the peaks in the power at $l\sim 6$ and $l\sim 11$ 
which correspond to the first and the second ISW ridge give
a bad fit to the COBE data for $L\sim 1.7H_o=0.17\times 2R_\ast$.
However, the parameter range which gives a good fit to the data 
is wider for the low-matter-density model. 
It should be emphasised that for both flat models
there is a parameter region in 
which the fit is much better than the infinite counterpart.
For example, the peaks at $l\sim 4$ and $l\sim 9$ in the angular 
power of the low-matter-density model with 
$L\sim 2.8H_0^{-1}$ give a much better fit to the COBE data
than the infinite counterpart. 
In fact the quadrapole component in the COBE data is very low 
and the angular power is peaked at $l\sim 4$. 
For infinite flat $\Lambda$-models with
a scale-invariant initial spectrum such features
can be a problem since the ISW contribution gives an excess power
on large angular scales.
\BF
\centerline{\psfig{figure=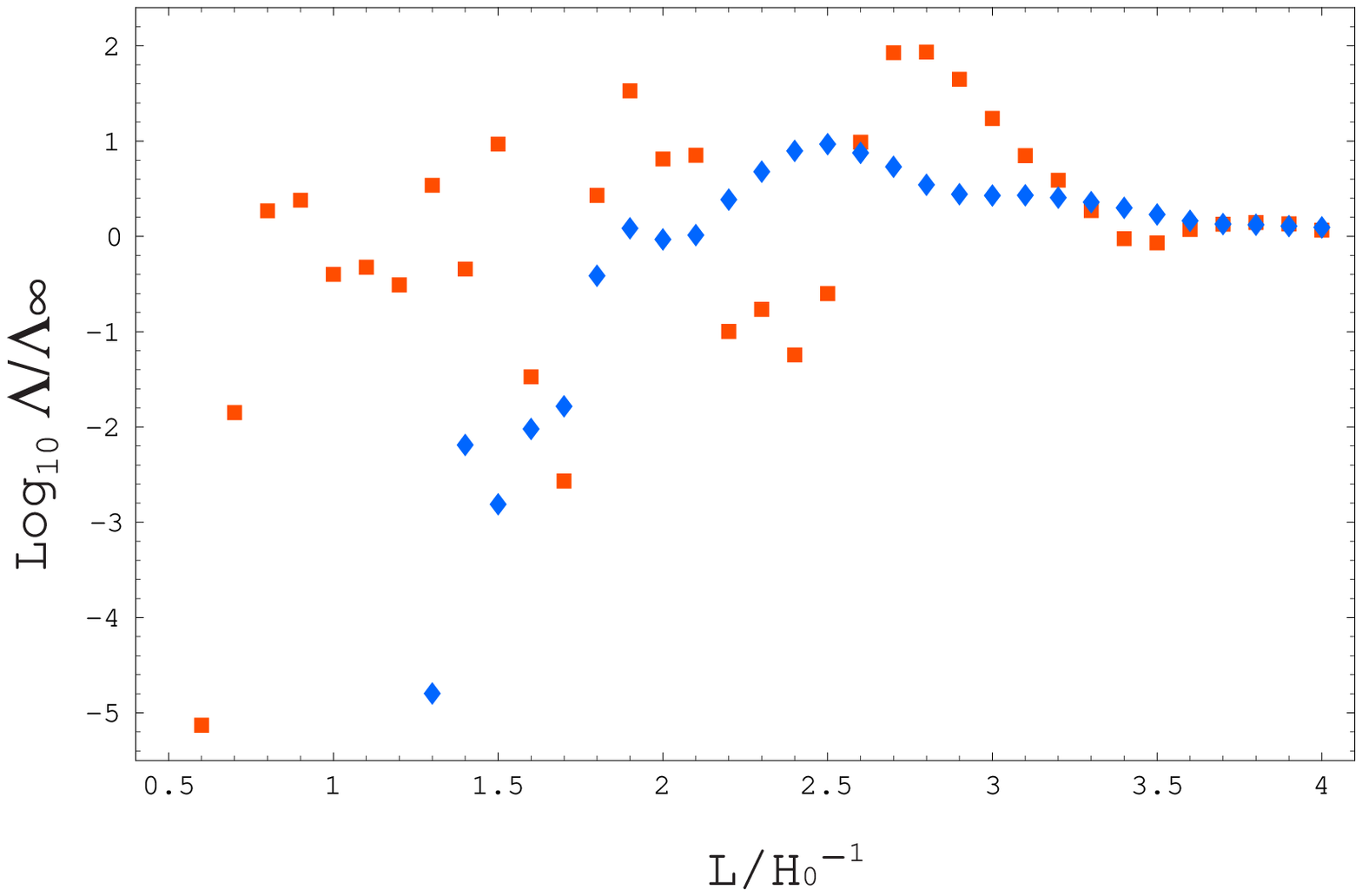,width=13cm}}
\caption{The likelihoods of the 3-torus models 
with $(\Omega_m,\Omega_\Lambda)=(1.0,0)$ (diamond) 
and (0.1,0.9) (box)
relative to the infinite models with the same density parameters
are plotted. 
The likelihoods are marginalised over the quadrapole normalisation
$Q\equiv(5C_2/(4 \pi))^{1/2}$. An isotropic Gaussian approximation has 
been used for computing the likelihoods. The COBE-DMR data is
compressed to 924 pixels at ``resolution 5''.  
}
\label{fig:LLres5F}
\EF
Next, we carry out a full Bayesian analysis 
in which all the elements of $<a_{lm}a_{l'm'}>$ are included.
Because of the limit in the CPU power we further compress
the data at ``resolution 5'' pixels to ``resolution 3''$(20.4^o)^2$
pixels in galactic coordinates for which there are 60 pixels surviving the 
extended galactic cut. Although the information on smaller angular scales 
$l>10$ is lost, we expect that they still provide us a sufficient 
information for discriminating the effect of the non-trivial topology
since it is manifest on large angular scales.
The computation has been done in a similar manner as previous 
analysis except for the covariance matrix for which we use (\ref{eq:M})
instead of (\ref{eq:MI}). 
The likelihoods $\Lambda$ are computed for a 
total of 2000 random orientations 
for each model using a vector parallel supercomputer vpp 800. 
The approximated likelihoods which depend on only 
the power spectra are also computed for comparison. 
\BF
\centerline{\psfig{figure=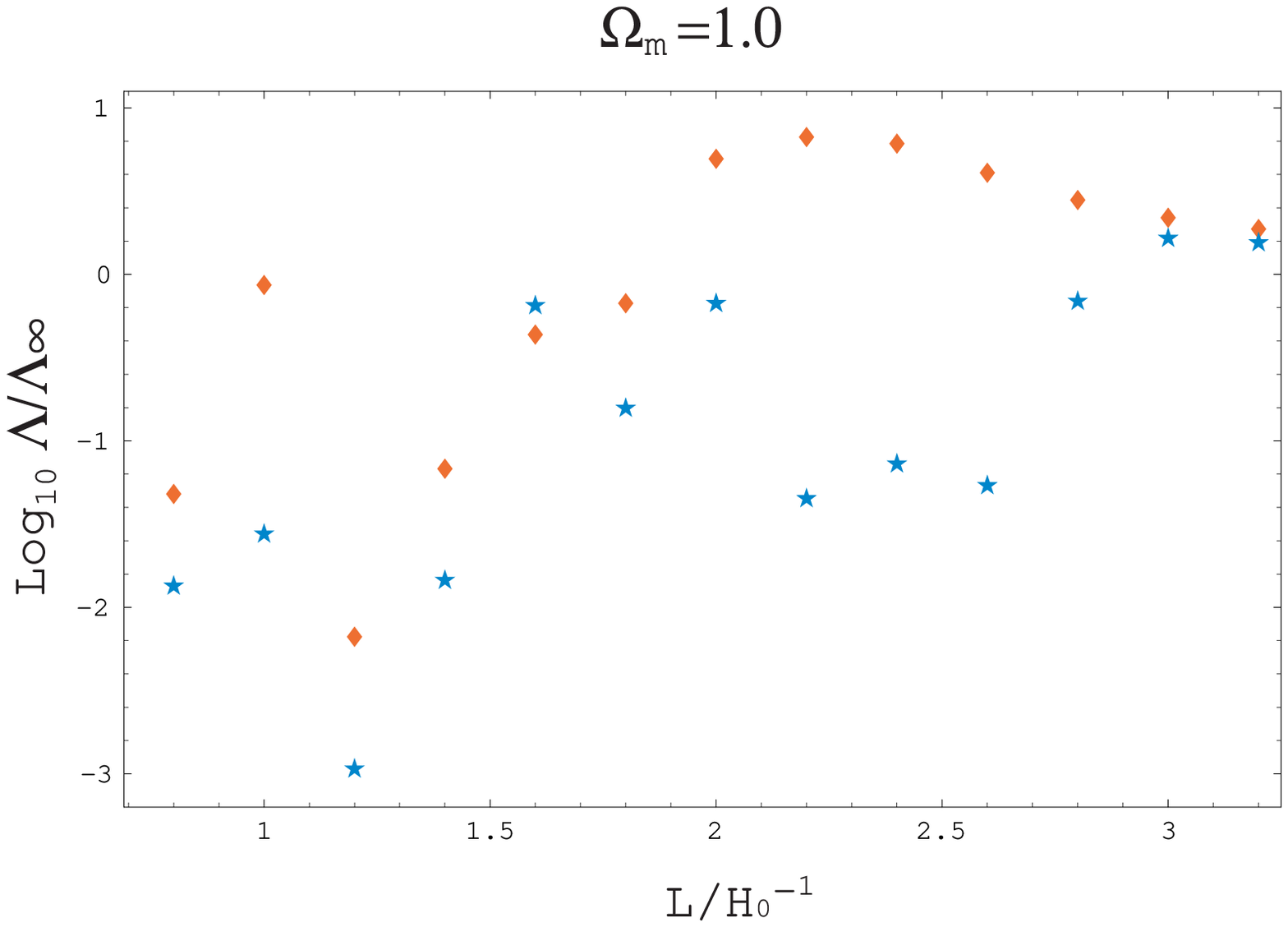,width=8.8cm}
\psfig{figure=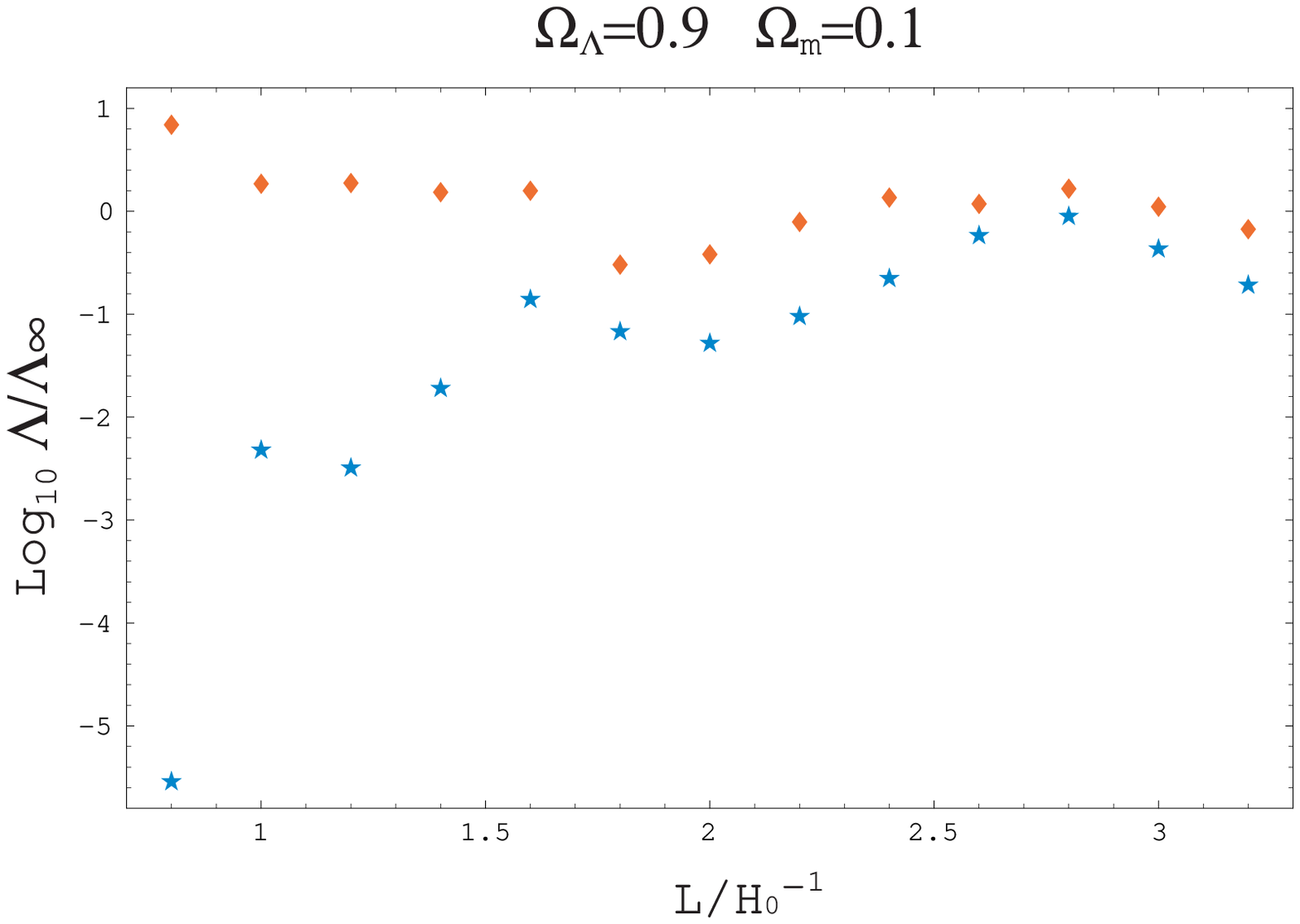,width=9cm}}
\caption{The likelihoods of the 3-torus models 
relative to the infinite models with the same density parameters
marginalised over 2000 orientations (star) and that using only
the power spectrum (diamond). 
The inverse-noise-variance-weighted
average map of the COBE-DMR data is compressed to 60 pixels
(resolution 3).    
All the likelihoods are also marginalised over the quadrapole normalisation
$Q\equiv(5C_2/(4 \pi))^{1/2}$. 
}
\label{fig:LLres3F}
\EF
As shown in figure \ref{fig:LLres3F}, the discrepancy between the 
likelihood marginalised over the orientation of the observer 
and the approximated likelihood is prominent for models with small volume
in which the effect of the non-trivial topology is significant. 
Let us estimate the size of the cell 
for which the effect becomes insignificant.
For the ``standard'' model with $\Omega_m\!=\!1.0$, the volume of
the observable region is $4 \pi R_\ast^3/3\!\sim\! 34$ which gives 
the critical scale $L_c\!\sim\!3.2 H_0^{-1}$ for which the volume of the
cell is comparable to the volume of the observable region at present.  
For the low-matter-density model 
a significant amount of large-angle fluctuations are produced 
at the $\Lambda$-dominant epoch $z\!\sim\!1$. 
Therefore one should compare the volume of the 
sphere with radius $\eta(z\!=\! 0)-\eta(z\!=\!1)\!\sim\! 0.9 H_0^{-1}$
to the volume of the cell instead of the observable region
which gives $L_c\!\sim\!1.5 H_0^{-1}$.   
We can see from figure \ref{fig:LLres3F} that these 
estimates well agree with the numerical result. 
\\
\indent
We can give two explanations for the discrepancy 
although they are related each other.
One is the non-Gaussianity in the fluctuations 
and another one is the correlations between $a_{l m}$'s 
owing to the global anisotropy in the background geometry.
\\
\indent
Suppose a random variable $Z=XY$ in 
which $X$ obeys a distribution function $E(X)$ which is even and
$Y$ obeys a distribution function $F(Y)$. Then the  
distribution function G of $Z$ is given by 
\BE
G(Z)=\int \f{E(Z/Y)}{|Y|}F(Y) dY,
\EE
which is apparently even. 
Because the fluctuations are written in terms of a sum of 
products of Gaussian variable  $\Phi(k)$ (with zero average) 
times a non-Gaussian variable $Y_{lm}(\hat \k)$, the skewness
in the distribution function of $a_{lm}$ marginalised over the
orientation is zero although the kurtosis is non-zero. 
\\
\indent
The correlations in $a_{lm}$'s are the consequence 
of the gap between the degree of freedom of $(\hat \k)$
and $(l,m)$. The degeneracy number in a $k$-mode 
(=the number of the direction $\hat \k$) is 
much less than the number of relevant ``quantum numbers''
$(l,m)$\footnote{Here we assumed that for each degenerated 
mode specified by $\hat \k$, $\Phi(k)$  
are independent Gaussian variables.} if the scale
(=$2\pi/k$) is comparable 
to the length of the side $L$.
Taking an ensemble average over the initial condition we have
\BE
<a_{lm}a_{l'm'}>=\sum_{\k \ne 0} \f{8 \pi^3}{k^3L^3} {\cal{P}}_\Phi(k)
Y_{lm}(\hat \k)Y_{l'm'}(\hat \k)
F_{k l}F_{k l'},
\label{eq:aa}
\EE
where $F_{kl}$ is given by (\ref{eq:psT3}). The sum does not
vanish in general when $(l,m)\ne(l',m')$ which is the consequence
of the global anisotropy in the background geometry. 
However, if one takes an average of
(\ref{eq:aa}) over $\hat \k$, one finds that  
all off-diagonal elements vanish because of the orthogonality 
of the spherical harmonics.
Similarly one can consider 4-point correlations
$<a_{l_1 m_1}a_{l_2 m_2}a_{l_3 m_3}a_{l_4 m_4}>$. In this case,
all the off-diagonal elements $(l_i,m_i)\ne(l_j,m_j)$ do not
necessarily vanish even if one takes an average over $\hat\k$.  
Thus the non-Gaussianity for the flat 3-torus models 
contrasts sharply with the one for the compact hyperbolic models
in which the pseudo-random Gaussian property of the 
expansion coefficients $b_{klm}$ (which are obtained by
expanding the eigenmode in terms of eigenmodes in the universal covering 
space)\cite{Inoue1,Inoue3} renders off-diagonal elements always
vanish if an average is taken over the position of the
observer(although the kurtosis is non-zero). 
The difference can be attributed to the property of eigenmodes.
In a less rigorous manner the property of the eigenmodes 
which are projected onto a sphere with large radius 
can be stated as follows:
\ti{Projected eigenmodes of compact hyperbolic spaces 
are ``chaotic'' whereas those of compact flat spaces  
are ``regular''}. 
\\
\indent
As for the constraint on the size of the cell, 
we set a slightly severe condition for the 
low-matter-density model since the likelihood of the infinite counterpart
is $\sim 10^{-1}$ of that of the infinite ``standard'' model due
to a slight boost on the large angular scales caused by the 
ISW effect.  Together with the previous analysis using the data on 
the ``resolution 5'' pixels, 
the conditions on the relative likelihood 
$\log_{10}(\Lambda/\Lambda_\infty)\!>\!-2$
and $\log_{10}(\Lambda/\Lambda_\infty)\!>\!-1$ yield
$L\ge 1.6H_0^{-1}$ and $L\ge2.2 H_0^{-1}$ 
for the ``standard'' 3-torus model $(\Omega_m,\Omega_\Lambda)=(1.0,0)$ 
and the low-matter-density 3-torus model 
$(\Omega_m,\Omega_\Lambda)=(0.1,0.9)$,
respectively. Here $\Lambda_\infty$ denotes
the likelihood of the infinite counterpart with the same 
density parameters. The maximum number $N$ of images of the cell  
within the observable region at present is 8 and 49 for the
former and the latter, respectively. 
Note that the constraint 
on the ``standard'' 3-torus model is consistent with
the previous result\cite{Stevens,Oliveira1}.
\pagebreak
%%%%%%%%%%%%%%%% SECTION V %%%%%%%%%%%%%%%%%%%
\section{Summary}
In this paper the CMB anisotropy in a flat 3-torus model with
or without the cosmological constant has been investigated. 
Using the COBE-DMR data,
we have done a full Bayesian analysis incorporating the effect of
anisotropic correlation. It has turned out that the
constraint on the low-matter-density model  
is less stringent compared to the ``standard'' model 
with $\Omega_m\!=\!1.0$. The reason is that the large-angle
fluctuations (on COBE scales) are produced at late time
well after the last scattering.
The physical size of these fluctuations is of the order
of the size of the cell or less. Hence the effect of the 
non-trivial topology becomes insignificant. We expect that 
the result does not significantly change for
other compact flat models with different topology\cite{Levin1}
or with different shape of the cell since the physical effect 
does not change. 
\\
\indent
We have seen that the analysis using the angular power spectrum 
is not enough since the background geometry is globally anisotropic.
Even if the power spectrum well agrees with the data, it does not 
guarantee the validity of the model since the power spectrum has 
information of only isotropic components in the 2-point correlations.
If the background geometry is globally anisotropic then 
the fluctuations form an anisotropic 
Gaussian field for a given orientation. 
On the other hand, the fluctuations become non-Gaussian
if one marginalises the distribution function over the
orientation\cite{Magueijo}. 
For locally Friedmann-Robertson-Walker models which are spatially
compact, we expect to observe zero skewness but 
non-zero kurtosis provided that 
the initial fluctuation is Gaussian. 
\\
\indent
However, for low-matter density models, anisotropy 
in the statistically averaged
correlation on large angular scale cannot be so large 
since the physical size of 
observed temperature fluctuations that has been prodeced well after
the last scatteing is much smaller than the actual 
size of the space(i.e. topological
identification scale). 
\\
\indent
In the case of flat topology, the discrete eigenmodes
have ``regular'' features which lead to significant 
correlations in $a_{lm}$'s. Even if marginalised over
the orientation, the correlations do not completely disappear.   
This contrasts with the case of hyperbolic topology in which
the discrete eigenmodes are ``chaotic'' and
the correlations in  $a_{lm}$'s disappear
if one takes an average over the position of the observer.  
\\
\indent
In order that we would be able to observe 
the periodic structure in the fluctuations for the 3-torus models,
we need to have  $L/(2 R_\ast)<1$ 
where $R_\ast$ is the comoving radius of the 
last scattering surface. Because the obtained constraints
give $L/(2 R_\ast)>0.40,0.22$ for $(\Omega_m,\Omega_\Lambda)=(1.0,0)
,(0.1,0.9)$, respectively, 
we still have a great chance of the first detection of
the non-trivial topology of the universe by the future satellite missions 
such as MAP and \ti{Planck} which can survey the CMB in the full sky
with high resolution by measuring some specific signatures
(e.g. the circles test \cite{Cornish2} or the non-Gaussianity 
\cite{Inoue3}). 
%%%%%%%%%%%%%%%%%%ACKNOWLEDGMENTS%%%%%%%%%%%%%%%%%%%%%%%
\section*{Acknowledgments}
I would like to thank N. Sugiyama, T. Chiba for their useful
advice and comments and A.J. Bandy for his advice 
on the use of the COBE data.
The numerical computation in this work was carried out  
at the Data Processing Center in Kyoto University and 
Yukawa Institute Computer Facility. 
K.T. Inoue is supported by JSPS Research Fellowships 
for Young Scientists, and this work is supported partially by 
Grant-in-Aid for Scientific Research Fund (No.9809834). 
%======================================%
%<<<<<<<<<<<< Bibliography >>>>>>>>>>>>%
%======================================% 

\end{document}